 \documentclass[twocolumn,showpacs,amsmath,prd,amssymb,showkeys,superscriptaddress,aps,longbibliography,nofootinbib]{revtex4-2}

\pdfoutput=1

\usepackage{graphicx}% Include figure files
\usepackage{dcolumn}% Align table columns on decimal point
\usepackage{mathtools}
\usepackage{scalerel}
\usepackage{bm}% bold math
\usepackage{upgreek}
\usepackage{mathrsfs}
\usepackage{float}
\usepackage[caption=false]{subfig}
\usepackage{array}
\usepackage{verbatim}
\usepackage{amstext}
\usepackage{amsmath}
\usepackage{amssymb}
\usepackage{slashed}
\usepackage[bbgreekl]{mathbbol}
\usepackage{indentfirst}
\usepackage{xspace}
\usepackage[T1]{fontenc}
\usepackage[latin9]{inputenc}
\usepackage{color}
\PassOptionsToPackage{normalem}{ulem}
\usepackage{ulem}
\usepackage[unicode=true,pdfusetitle,
  bookmarks=true,bookmarksnumbered=false,bookmarksopen=false,
  breaklinks=false,pdfborder={0 0 1},backref=false,colorlinks=true]
  {hyperref}
\hypersetup{
  citecolor=blue}

\newcommand{\delcp}         {\delta_{cp}}

\newcommand{\ax} {a^{X}_{\alpha \beta}}
\newcommand{\ay}  {a^{Y}_{\alpha \beta}}
\newcommand{\ctx}  {c^{TX}_{\alpha \beta}}
\newcommand{\cty} {c^{TY}_{\alpha \beta}}
\newcommand{\cxz}  {c^{XZ}_{\alpha \beta}}
\newcommand{\cyz} {c^{YZ}_{\alpha \beta}}
\newcommand{\cxx}  {c^{XX}_{\alpha \beta}}

\newcommand{\cxy} {c^{XY}_{\alpha \beta}}

\newcommand{\emu} {e \mu}
\newcommand{\et} {e \tau}
\newcommand{\mt} {\mu \tau}

\makeatother

\begin{document}

\title{Exploring Non-Isotropic Lorentz Invariance Violation Through Sidereal Effect at DUNE}

\newcommand{\cusb}{Department of Physics,
Central University of South Bihar, Gaya 824236, India}

\newcommand{\bhu}{Department of Physics, Institute of Science,
Banaras Hindu University,
Varanasi 221005, India.}

\newcommand{\corrsks}{saurabhshukla@cusb.ac.in}
\newcommand{\corrls}{lakhwinder@cusb.ac.in}

\author{ Shashank~Mishra} \affiliation{ \cusb }\affiliation{ \bhu }
\author{ Saurabh~Shukla} \altaffiliation{ \corrsks }\affiliation{ \cusb }\affiliation{ \bhu }

\author{ Lakhwinder~Singh }  \altaffiliation{\corrls}\affiliation{ \cusb }
\author{ Venktesh~Singh }   \affiliation{ \cusb }

\date{\today}

\begin{abstract}
Lorentz Invariance Violation (LIV) presents an intriguing opportunity to investigate fundamental symmetries, with neutrinos serving as a particularly effective probe for this phenomenon. Long-baseline neutrino experiments, such as the Deep Underground Neutrino Experiment (DUNE), excel at exploring non-isotropic LIV, especially through the observation of sidereal effects. This study comprehensively examines the full parameter space of non-isotropic, non-diagonal LIV parameters with sidereal dependence, focusing on two distinct flux scenarios: a low-energy flux and a tau-optimized flux. Through this analysis, we derive more stringent constraints on LIV parameters. Our results indicate that DUNE may achieve enhanced sensitivity for some LIV parameters, exceeding all previously established limits and marking a significant advancement in the investigation of LIV.

   \end{abstract}

\maketitle

\section{\label{sec:level1}Introduction:\protect}
Neutrinos offer a promising avenue for investigating new physics beyond the Standard Model (BSM) for both theoretical and experimental physicists. The first compelling evidence for BSM is the neutrino mixing and non-zero masses, suggested by several neutrino oscillation experiments with different sources. The Lorentz and CPT symmetry breaking are predicted phenomenon at the Planck scale that doesn't necessitate changing either general theory of relativity or quantum field theory~\cite{Kostelecky:1988zi, Kostelecky:1991ak,Gambini:1998it,CliffP.Burgess_2002}. It is no surprise that neutrino oscillation experiments hold the discovery potential for experimental evidence of Lorentz and CPT violation~\cite{Diaz:2009qk,Kostelecky:2003xn,Kostelecky:2004hg}. The Standard Model Extension (SME) framework provides an effective field theory for describing low-energy effects at the Planck scale, incorporating all possible Lorentz Invariance Violation (LIV) operators, along with the full spectrum of particles and Standard Model interactions~\cite{Colladay:1998fq, Bluhm:2005uj,RevModPhys.83.11}. Thus, the SME is a viable framework for studying LIV effects.

 Both isotropic and non-isotropic forms of LIV affects neutrino oscillations in distinct ways. Isotropic LIV parameters modifiy the overall oscillation probability in a direction-independent way, leading to deviations from the conventional L/E dependence. On the other hand,  non-isotropic LIV parameters introduce directional dependence in the oscillation process, resulting in sidereal time-dependent oscillations due to the presence of a preferred direction in the Universe.

Long-baseline neutrino experiments are well-suited for studying non-isotropic LIV through the sidereal effect, due to the unique configuration of these experiments. Several neutrino experiments have investigated non-isotropic LIV parameters using sidereal effect, including LSND~\cite{LSND:2005oop}, MINOS-ND~\cite{MINOS:2010kat, MINOS:2012ozn}, MiniBooNE~\cite{MiniBooNE:2011pix}, IceCube~\cite{IceCube:2010fyu}, Double Chooz~\cite{DoubleChooz:2012eiq}, T2K~\cite{T2K:2017ega}, and DayaBay~\cite{DayaBay:2018fsh}. Furthermore, neutrinos have been employed in a range of phenomenological studies to investigate and elucidate the characteristics of LIV.~\cite{Delgadillo:2024bae, Cordero:2023hua, Barenboim:2018ctx, Barenboim:2023krl, Cordero:2024hjr, Agarwalla:2023wft, Crivellin:2020oov}

In this article, we present the sensitivity to LIV parameters in non-isotropic time dependence scenarios for the Deep Underground Neutrino Experiment (DUNE) ~\cite{DUNE:2020lwj}.
We systematically investigate the impact of introducing each non-zero LIV parameter individually, analyzing the resulting changes in oscillation probabilities, and evaluate the sensitivity of the DUNE to constrain non-isotropic LIV parameters. Given that several ongoing studies aim to understand the implications of various flux scenarios for DUNE's physics potential~\cite{Giarnetti:2024mdt, Siyeon:2024pte, Masud:2017bcf, Rout:2020cxi}, we consider two publicly available flux scenarios: low energy (LE) and high energy (HE) ~\cite{DUNE:2020ypp}, to investigate the experiment's sensitivity to beam energy.

The organization  of this article is as follows. In Sec.~\ref{sect::formulation}, we give
the general formulation of LIV considered in the present work. In Sec.~\ref{sec:simulation}, we outline the neutrino beam parameter considered in the present work in the context of DUNE. In Sec.~\ref{sec::resDis}, we present and discuss our sensitivity of DUNE for different senarios. The summary is in Sec.~\ref{sec::summary}.

\section{Formalism}
\label{sect::formulation}

The leading-order effective Lagrangian density for the neutrinos and antineutrinos with Lorentz and CPT violation can be written as~\cite{PhysRevD.69.016005,PhysRevD.85.096005}:  
\begin{equation}
    \mathcal{L}=\frac{1}{2}\bar{\psi}(i\partial-M-\hat{\mathcal{Q}})\psi + h.c.
\label{lagliv}
\end{equation}
where the spinors $\psi$ and $\bar{\psi}$ represent the fields associated with the neutrino and antineutrino, respectively. The first term of the Lagrangian generates the kinetic term, while the second term introduce the arbitrary mass matrix, and the third term captures the effect of Lorentz invariance violation, represented by a generic Lorentz invariance violating operator ($\hat{\mathcal{Q}}$). When considering renormalizable Dirac couplings into account, the Lorentz invariance violating portion of the Lagrangian can be expressed as follows~\cite{PhysRevD.85.096005}:

\begin{equation}
    \begin{aligned}[b]
        & \mathcal{L}_{\rm LIV} = -\frac{1}{2}\left[a^{\mu}_{\alpha\beta} \bar{\psi_{\alpha}}\gamma_{\mu}\psi_{\beta}+ b^{\mu}_{\alpha\beta} \bar{\psi_{\alpha}}\gamma_{5}\gamma_{\mu}\psi_{\beta}\right]\\
        & -\frac{1}{2}\left[-ic^{\mu\nu}_{\alpha\beta} \bar{\psi_{\alpha}}\gamma_{\mu}\partial_{\nu}\psi_{\beta} - id^{\mu\nu}_{\alpha\beta} \bar{\psi_{\alpha}}\gamma_{5}\gamma_{\mu}\partial_{\nu}\psi_{\beta} \right] + h.c
    \end{aligned}
\end{equation}
where $a^{\mu}_{\alpha\beta}$, $b^{\mu}_{\alpha\beta}$, $c^{\mu\nu}_{\alpha\beta}$ and $d^{\mu\nu}_{\alpha\beta}$ are the Lorentz symmetry breaking parameters defined in the flavor basis. The first two terms  including $a^{\mu}_{\alpha\beta}$ and  $b^{\mu}_{\alpha\beta}$ parameters are CPT-violating LIV, while the third and the fourth terms ($c^{\mu\nu}_{\alpha\beta}$ and $d^{\mu\nu}_{\alpha\beta}$)  are CPT-conserving LIV term.
When considering only the interactions of left-handed neutrinos, these quantities can be parameterized according to the framework outlined in ~\cite{PhysRevD.85.096005} as, 
\begin{equation}
(a_{L})^{\mu}_{\alpha\beta}=(a + b )^{\mu}_{\alpha\beta}, \hfill ~~~  (c_{L})^{\mu\nu}_{\alpha\beta}=(c  + d )^{\mu\nu}_{\alpha\beta}.
\end{equation}

Thus, $(a_{L})^{\mu}_{\alpha\beta}$ and $(c_{L})^{\mu\nu}_{\alpha\beta}$ are two hermitian matrices encoded 
the LIV effect in the interaction hamiltonian. These hermitian matrices modify the standard neutrino oscillation Hamiltionian, by adding a new term $H_{LIV}$ to the standard interaction term which have vacuum and matter contributions.
\begin{equation}
H = H_{SI} + H_{LIV},
\label{Htotal}
\end{equation}

The standard Hamiltonian for neutrino oscillations in matter can be expressed as, 
\begin{equation}\label{eq:Hm}
H_{SI} = \frac{1}{2E} \left[ U \, \text{diag}(0, \Delta m_{21}^2, \Delta m_{31}^2) \, U^\dagger \right] + V_m, 
\end{equation}
where neutrino mixing of  flavor states is described by the matrix $U$, which includes a CP-violating phase $\delta_{CP}$ and three mixing angles, $\theta_{ij}$.
The neutrino energy is denoted by $E$, and the two established mass square discrepancies are represented by $\Delta m_{21}^2$ and $\Delta m_{31}^2$. The matter potential in a long-baseline experiment is accounted by $V_m$. In the case of Lorentz invariance violation, the perturbation in the effective Hamiltonian for neutrino oscillations within the framework of the SME can be expressed as follows~\cite{Kostelecky:2003xn}:

\begin{equation}
  ((\mathcal{H}_{liv})_{\nu\nu})_{\alpha\beta}  =  \frac{1}{|\vec{p}|}[(a)^{\mu} p_{\mu}  - (c)^{\mu\nu} p_{\mu} p_{\nu}]_{\alpha \beta},
  \label{hLIV}
\end{equation}
where $(a)^{\mu}$ and $(c)^{\mu\nu}$ are $3 \times 3$ complex matrices, referred to as coefficients, with their spacetime indices $\mu, \nu$ encoding the nature of the symmetry breaking. If only the time components of these coefficients are nonzero, it signifies isotropic (direction-independent) Lorentz violation. On the other hand, nonzero spatial components introduce anisotropy, causing direction-dependent effects on neutrino behavior. The neutrino four-momentum is given by 
$p_{\mu} = (E; \vec{p})$, where  $\vec{p}$ represents the direction of neutrino propagation.  The CP conjugate of the Hamiltonian given in Eq~\ref{hLIV} describes right-handed antineutrinos, which retains the same form as the original Hamiltonian but with $(a)^{\mu} = -(a)^{\mu}$ and $(c)^{\mu\nu} = (c)^{\mu\nu}$.

The observation of spacetime anisotropy would serve as smoking-gun signature of Lorentz invariance violation, stemming from preferred directions defined by the coefficients $(a)^{\mu}$ and $(c)^{\mu\nu}$. These Lorentz violation coefficients act as fixed background fields that influence the structure of spacetime. To detect such violations, one can investigate how these background fields interact with the direction of neutrino propagation. By measuring neutrino beams in different orientations, any variations in the data may point to the presence of spacetime anisotropy.

An Earth-based neutrino experiment where both the source and detector are fixed on the Earth's surface, proivde natural interferometer with senstivity compare to prcecsie optical experiments. The rotation of the Earth around its axis introduces a sidereal variation in neutrino oscillation probabilities. This variation occurs because the orientation of the experiment changes relative to any preferred directions in space, potentially associated with Lorentz invariance violation. As a result, oscillation probabilities can fluctuate at multiples of the Earth's sidereal frequency, $\omega_{\oplus} \approx 2\pi / (23 \text{ h } 56 \text{ min})$, which is the rate at which the Earth completes one full rotation relative to distant stars.

The effects of Lorentz violation are preserved across different reference frames, therefore, there is no preferred reference frame for measurements. In order to compare results from different experiments in a meaningful way, it's essential to use a common reference frame. To achieve this, a universal coordinate system is necessary, which should be reasonably inertial over the timescale of the experiments. This universal coordinate system allows for a standardized comparison of experimental results, ensuring that all measurements and violations are interpreted consistently across various setups. The Sun-centered celestial equatorial frame, with coordinates (T,X,Y,Z), is commonly used to report experimental results, providing a consistent and systematic approach to detecting Lorentz-violating effects. In this frame, the Earth's rotation axis aligns with its orbital axis, tilting the orbital plane by 23.4 degrees, which defines the Z-axis. The X-axis points toward the vernal equinox, while the Y-axis completes the right-handed coordinate system, ensuring an accurate and uniform reference for measuring potential Lorentz-violating signals across experiments.

The effective Hamiltonian, which incorporates sidereal time dependencies in the Sun-centered celestial equatorial frame, can be reorganized to highlight the dependence of its parameters on sidereal frequency as follows~\cite{PhysRevD.109.075042}:
\setlength\abovedisplayskip{5pt}
\begin{widetext}
\begin{equation}
  \begin{split}
       (\mathcal{H}_{LIV})_{\alpha\beta} = (C)_{\alpha\beta} + R [ a_{\alpha\beta}^{X} - 2 E (c^{TX})_{\alpha\beta} + 2 E N_{z} (c^{XZ})_{\alpha\beta}] sin(\omega_{\oplus} T - \Phi_{orientation})~- \\
   R [ a_{\alpha\beta}^{Y} - 2 E (c^{TY})_{\alpha\beta} + 2 E N_{z} (c^{YZ})_{\alpha\beta}] cos(\omega_{\oplus} T - \Phi_{orientation})~+ \\
   R^{2} [ E \frac{1}{2}((c^{XX})_{\alpha\beta} - (c^{YY})_{\alpha\beta} ) ] cos(2 (\omega_{\oplus} T - \Phi_{orientation}))~+ \\
   R^{2} [ E (c^{XY})_{\alpha\beta}] sin(2 (\omega_{\oplus} T - \Phi_{orientation})),
  \end{split}
  \label{hLIVexpanded}
\end{equation}
\end{widetext}
In the Sun-centered frame, $T$ represents the local sidereal time, describing how the Earth rotates relative to a sidereal star. The factors related to the beam and detector, $\Phi_{o}$ and $R$, can be expressed in terms of the directional components $N^{X}$ and $N^{Y}$.
\begin{equation}
\begin{split}
  \Phi_{o} = \tan^{-1}(N^{X}/N^{Y}),\\
  R = \sqrt{ (N^{X})^2 + (N^{Y})^2 }., 
  \end{split}
\end{equation}

The directional factors ($N^{X}$, $N^{Y}$, and $N^{Z}$) in the LIV Hamiltonian play a critical role in characterising the spatial orientation of the experimental setup with respect to the Earth's surface. These variables offer essential characteristics for examining directional dependencies in many studies related to physics. The directional factors are represented by the following: the colatitude of the detector ($\chi$); the angle between the beam and the south measured towards the east ($\phi$) known as the "bearing"' angle; and the angle between the beam and the vertically upward direction ($\theta$) known as the "Zenith"' angle~\cite{Kostelecky:2004hg}. 
\begin{equation}
  \begin{split}
  N^{X} &= \cos\chi\sin\theta\cos\phi + \sin\chi\cos\theta,\\
  N^{Y} &= \sin\theta\sin\phi,\\
  N^{Z} &= -\sin\chi\sin\theta\cos\phi + \cos\chi\cos\theta, 
  \end{split}
  \label{orientationbeam}
\end{equation}

In the LIV Hamiltonian given in Eq.~\ref{hLIVexpanded}, the term $(C)_{\alpha\beta}$ does not exhibit any sidereal dependency. The parameters $(a)^{X}_{\alpha \beta}$, $(a)^{Y}_{\alpha \beta}$, $(c)^{TX}_{\alpha \beta}$, $(c)^{TY}_{\alpha \beta}$, $(c)^{XX}_{\alpha \beta}$, $(c)^{XY}_{\alpha \beta}$, $(c)^{XZ}_{\alpha \beta}$, $(c)^{YY}_{\alpha \beta}$, and $(c)^{YZ}_{\alpha \beta}$ are responsible for the sidereal modulation of the perturbed Hamiltonian terms.

The nature of the LIV perturbation Hamiltonian, as shown in Eq.~\ref{hLIVexpanded}, which has X-type ($\ax$, $\ctx$, $\cxz$)  parameters appearing with a sinusoidal term and Y-type ($\ay$, $\cty$, $\cyz$)parameters appearing with a cosine term, can be introudce approximately 6 sidereal hours phase difference~\cite{PhysRevD.109.075042}. The parameters $(a)^{X}_{\alpha \beta}$, $(a)^{Y}_{\alpha \beta}$, $(c)^{TX}_{\alpha \beta}$, $(c)^{TY}_{\alpha \beta}$, $(c)^{XZ}_{\alpha \beta}$, and $(c)^{YZ}_{\alpha \beta}$ demonstrate sidereal modulation at half the sidereal frequency compared to the coefficients $(c)^{XX}_{\alpha \beta}$, $(c)^{XY}_{\alpha \beta}$, and $(c)^{YY}_{\alpha \beta}$.

If the contribution of LIV perturbation in Eq.~\ref{Htotal} is sufficiently small, the oscillation probabilities for both the appearance and disappearance channels can be expressed up to the leading order for the $\mu e$ and $\mu\mu$ channels, similarly as presented in ref ~\cite{Liao:2016hsa,Chaves:2018sih,Yasuda:2007jp,PhysRevD.109.075042}
    
\begin{widetext}
  \begin{equation}
  \begin{split}
     P_{\mu e}^{\rm LIV}  \simeq x^2 f^2 +2 x y fg \cos (\Delta+\delta_{CP})
    + y^2 g^2 + 4 r_A |{h}^{\rm{LIV}}_{e \mu}| 
    \big\{ xf \big [ f s_{23}^2 \cos (\phi^{\rm{LIV}}_{e \mu}+\delta_{CP})
      +g c_{23}^2 \cos( \Delta +\delta_{CP}+\phi^{\rm{LIV}}_{e \mu})\big] \\
     + yg \big[ g c_{23}^2 \cos \phi^{\rm{LIV}}_{e \mu} 
      + f s_{23}^2\cos (\Delta-\phi^{\rm{LIV}}_{e \mu})\big ]\big \}
    + 4 r_A |{h}^{\rm{LIV}}_{e\tau}| s_{23} c_{23} \big\{xf \big[ f \cos (\phi^{\rm{LIV}}_{e \tau}+\delta_{CP})
      - g\cos (\Delta +\delta_{CP}+\phi^{\rm{LIV}}_{e \tau}) \big] \\
       - yg[ g \cos \phi^{\rm{LIV}}_{e \tau}
      -f \cos(\Delta-\phi^{\rm{LIV}}_{e\tau})\big]\big\}+ 4 r_A^2 g^2 c_{23}^2 | c_{23} |{h}^{\rm{LIV}}_{e \mu}|
    -s_{23} |{h}^{\rm{LIV}}_{e\tau}||^2 + 4 r_A^2 f^2 s_{23}^2  | s_{23} |{h}^{\rm{LIV}}_{e \mu}|
    +c_{23} |{h}^{\rm{LIV}}_{e\tau}||^2 \\
     + 8 r_A^2 f g s_{23}c_{23} \big\{ c_{23} \cos \Delta \big[s_{23}(|{h}^{\rm{LIV}}_{e \mu}|^2 -|{h}^{\rm{LIV}}_{e \tau}|^2)
      + 2 c_{23} |{h}^{\rm{LIV}}_{e \mu}| |{h}^{\rm{LIV}}_{e \tau}|\cos(\phi^{\rm{LIV}}_{e \mu}-\phi^{\rm{LIV}}_{e \tau}) \big] \\
     - |{h}^{\rm{LIV}}_{e \mu}|| {h}^{\rm{LIV}}_{e \tau}| \cos (\Delta -\phi^{\rm{LIV}}_{e \mu}
    +\phi^{\rm{LIV}}_{e \tau})\big \} 
    +{\cal O}(s_{13}^2 a, s_{13}a^2, a^3),
    \label{pmue}
\end{split}
\end{equation}

\begin{equation}
  \begin{split}
    P_{\mu\mu}^{\rm LIV} & \simeq 1- \sin^2 2 \theta_{23}\sin ^2 \Delta  - |{h}^{\rm{LIV}}_{\mu\tau}|
    \cos \phi^{\rm{LIV}}_{{\mu\tau}} \sin 2 \theta_{23}
    \Big[ (2r_A\Delta )\sin^2 2\theta_{23}\sin 2\Delta + 4 \cos^2 2 \theta_{23}r_A\sin ^2\Delta\Big]\\
    & + (|{h}^{\rm{LIV}}_{\mu\mu}| - |{h}^{\rm{LIV}}_{\tau\tau}|)\sin^2  2 \theta_{23} \cos 2 \theta_{23}
    \Big[(r_A\Delta) \sin 2\Delta -2r_A \sin ^2 \Delta  \Big],
    \label{pmumu}
  \end{split}
\end{equation}
%\end{widetext}
where
\begin{equation}
  \begin{split}
    s_{ij}=\sin\theta_{ij},~~c_{ij}=\cos\theta_{ij},~~ 
    x=2s_{13}s_{23},~~ y=2rs_{12}c_{12}c_{23},~~ \Delta = \frac{\Delta m^2_{31} L}{4E},~~  r=|\Delta m^2_{21}/\Delta m^2_{31}|~~    \\
         r_A=\frac{2E}{{\Delta m}^2_{31}},~~
              f=\frac{\sin\big[\Delta (1-r_A(V_{CC}+{h}^{\rm{LIV}}_{ee}))  \big]}{1-r_A(V_{CC}+{h}^{\rm{LIV}}_{ee})},~~
              V_{CC}=\sqrt 2 G_F N_e,~~ 
    g=\frac{\sin\big[\Delta r_A(V_{CC}+{h}^{\rm{LIV}}_{ee})  \big]}{r_A(V_{CC}+{h}^{\rm{LIV}}_{ee})}.\\
    \hspace{0.5 true cm}\label{os-po}
  \end{split}
\end{equation}

  \end{widetext}
  
In the context of the LIV Hamiltonian, the matrix elements for antineutrinos are adjusted according to the changes outlined in the formalism section, where $V_{CC}$ changes to $-V_{CC}$ and $\delcp$ changes to $-\delcp$. To leading order, the appearance channels are affected by the $e\mu$ and $e\tau$ parameters, while the disappearance channels are influenced by the $\mu\tau$ parameters.

\section{Experimental and simulation details}
\label{sec:simulation}
The Deep Underground Neutrino Experiment (DUNE) is a multi-purpose detector designed to address some of the most fundamental questions in astrophysics and particle physics. At the forefront of its research, DUNE aims to explore critical topics such as neutrino mass hierarchy, the origins of the universe,  nucleon decay, and supernova burst detection etc. DUNE is also highly sensitive to atmospheric neutrinos, enhancing our understanding of cosmic events and particle interactions occurring in Earth's atmosphere. Search for the Lorentz and CPT violation to sidereal dependence is one of the major ancillary program of the DUNE~\cite{DUNE:2015lol}. The DUNE consists of two detectors exposed to a muon neutrino beam originating  at Fermilab. A high-precision near detector, located 575 m from the neutrino source ($41.829002^{\circ} N, 88.264039^{\circ} W $) on the Fermilab site, will be used to characterize the intensity and energy spectrum of this wide-band beam. A Far detector is located 1300 kilometers away at the Sanford Underground Research Facility in South Dakota ($44.351853^{\circ} N, 103.751452^{\circ} W$), which consists of four 10-kiloton liquid argon Time Projection Chambers (TPC).

\begin{figure}[!h]
         \includegraphics[height= 0.50\textwidth,width=0.5\textwidth]{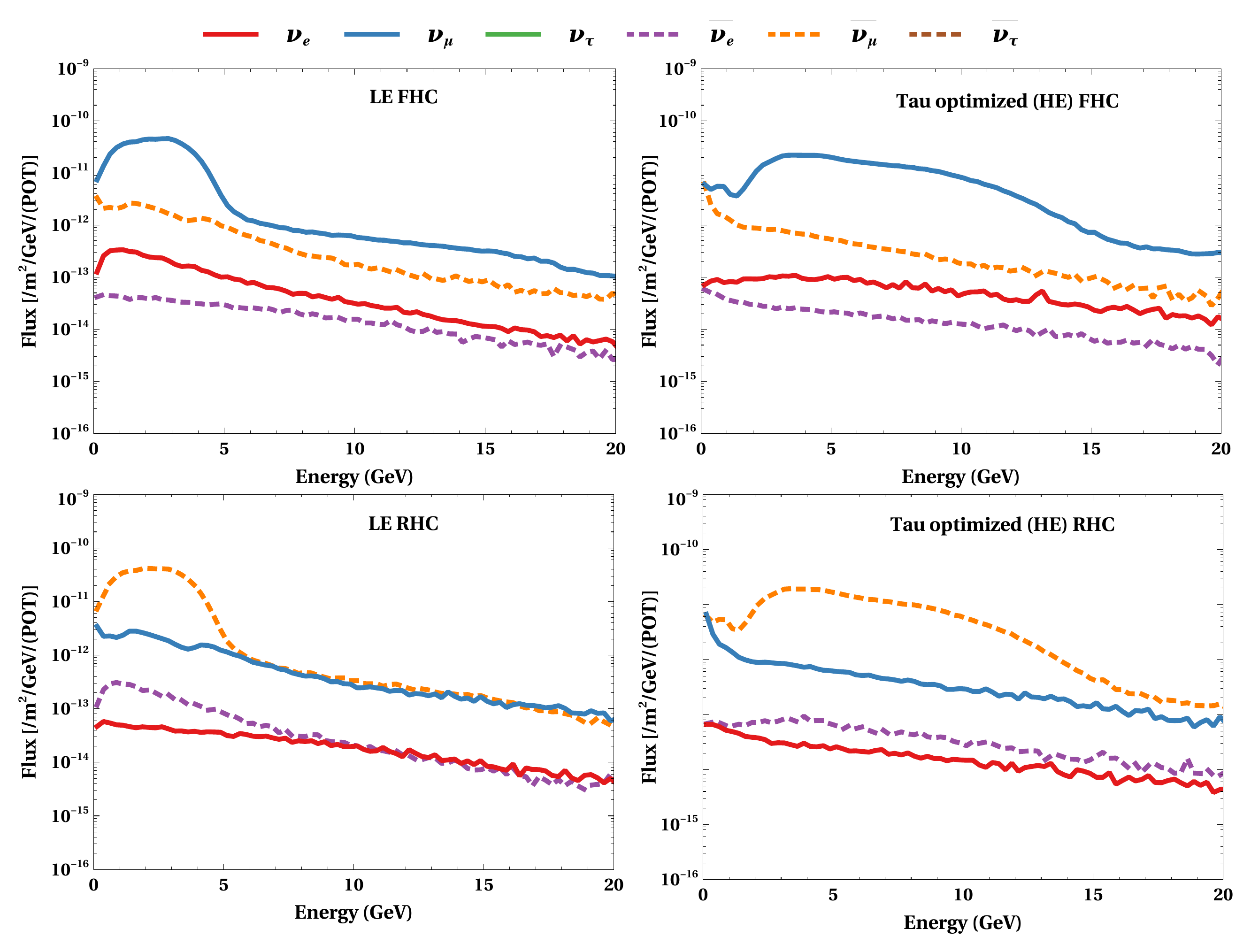}
         \caption{(Top) DUNE neutrino fluxes at the Far Detector (FD) for the FHC beam running are shown, with the low-energy (LE) flux depicted in the left panel and the tau-optimized flux in the right panel. (Bottom) Antineutrino flux at the FD for the RHC beam running is presented, with the LE flux on the left panel and the tau-optimized flux on the right panel~\cite{lammps}. }
      \label{dune_flux}
    \end{figure}

To simulate the neutrino oscillation for the long-baseline DUNE case, we adopted the GLoBES software~\cite{Huber:2004ka, Huber:2007ji} and plugin the most recent DUNE configuration files provided by
the collaboration~\cite{DUNE:2021cuw}. The sidereal effect from LIV parameter has been incorporated into the GLoBES simulation package. The true values for the standard neutrino oscillation parameters used in this work are as follows: $\theta_{12}$ = $33.48^{\circ}$, $\theta_{13}$ = $8.5^{\circ}$, $\theta_{23}$ = $45.0^{\circ}$, $\delta_{CP}$ = $195.0^{\circ}$, $\Delta m^{2}_{21}$ = $7.55 \times 10^{-5}$ $eV^{2}$, and $\Delta m^{2}_{31}$ = $2.50 \times 10^{-3} eV^{2}$. Simulations are carried out using two types of fluxes: the standard LE flux, which has a shorter energy range, and the tau optimized (HE) flux, which has an energy range of around 1 GeV to 10 GeV, as illustrated in figure~\ref{dune_flux}. The simulation runs in (5+5) years for (neutrino+anti-neutrino) modes, for each HE and LE fluxes.

\begin{table}[!h]
  \caption{DUNE FD orientation details used in the simulation.}
  \label{table3beamO}
   \setlength{\tabcolsep}{15pt}
 \renewcommand{\arraystretch}{1.5}
  \begin{tabular}{|c|c|}
    \hline 
    Parameter            & Value         \\ \hline \hline
    $\chi$   co-latitude  & $ 45.64814637^{\circ} $\\   \hline
    $\theta$ zenith angle   & $ 83.8^{\circ} $\\   \hline
    $\phi $  bearing       & $252.237^{\circ} $\\   \hline
      \end{tabular}
\end{table}
 
The DUNE far detector, located 1300 km from the neutrino source, offers a long baseline for studying neutrino behavior. DUNE functions as an on-axis experiment, providing a wide range of neutrino flux. The orientation specifics of the DUNE FD used in the simulation are presented in Table \ref{table3beamO}\footnote{For estimating the zenith and bearing angles, we have approximated the DUNE FD as being on the surface. This approximation is based on the fact that the baseline length (1300 km) is much larger than the underground depth (1.5 km) of the FD. The effect of accounting for the underground depth on the altitude and consequently on the zenith and bearing angles is minimal.}.

The dependency of non-isotropic LIV parameters  on directional parameters of experiment and neutrino energy is given in Table~\ref{tableM}. The DUNE experiment operates under conditions where $R \gg N_{z}$. Consequently, parameters such as $\cxy$ and $\cxx$, which scale with $R^2$, and $\ax$, which scales with $R$, exhibit stronger constraints in each flux scenario compared to $\cxz$, which scales with $R N_{z}$.

The dependence of non-isotropic Lorentz Invariance Violation (LIV) parameters on the directional parameters of the experiment and neutrino energy is summarized in Table~\ref{tableM}. Several experimental proposals aim to explore distinct physics scenarios using different flux types. Since Some LIV parameters exhibit energy dependence, it is natural to consider multiple flux scenarios in our current study. To investigate these effects, two beam configurations are employed, as referenced~\cite{DUNE:2020ypp}. Figure \ref{dune_flux} presents the DUNE flux profiles
for Low Energy (LE) flux in the left panel and the tau-optimized flux illustrated in the right panel. The reverse horn current (RHC) primarily corresponds to antineutrinos beam, while the forward horn current (FHC) mainly corresponds to neutrinos beam.

\begin{table}[h]
  \caption{Dependency of non-isotropic LIV parameter on directional parameters of Experiment and neutrino energy.}
  \label{tableM}
  \setlength{\tabcolsep}{12pt}
 \renewcommand{\arraystretch}{1.50}     
  \begin{center}
\begin{tabular}{|c|c|c|c|}
    \hline
    Parameter  & Energy   & Direction		& Total  \\ \hline
    $a^{X}$  & $E^{0}$  & R  		& $R$           	 \\ \hline
    $c^{TX}$ & E      & R 		& $2RE$        	\\ \hline
    $c^{XZ}$ & E      & $R N_{Z}$ & $2REN_{Z}$     \\ \hline
    $c^{XX}$ & E      & $R^{2}$ 		& $R^{2}E/2$ 	 \\ \hline
    $c^{XY}$ & E      & $R^{2}$  	& $R^{2}E$  	\\ \hline
\end{tabular}
 \end{center}
\end{table}

\section{Results and Discussion}
\label{sec::resDis}

\begin{figure*}[htbp]
  \begin{minipage}{0.48\textwidth}
     \centering
    \subfloat[Left: Appearance channel, Right: Disappearance channel in the SM case]{\includegraphics[width=\textwidth]{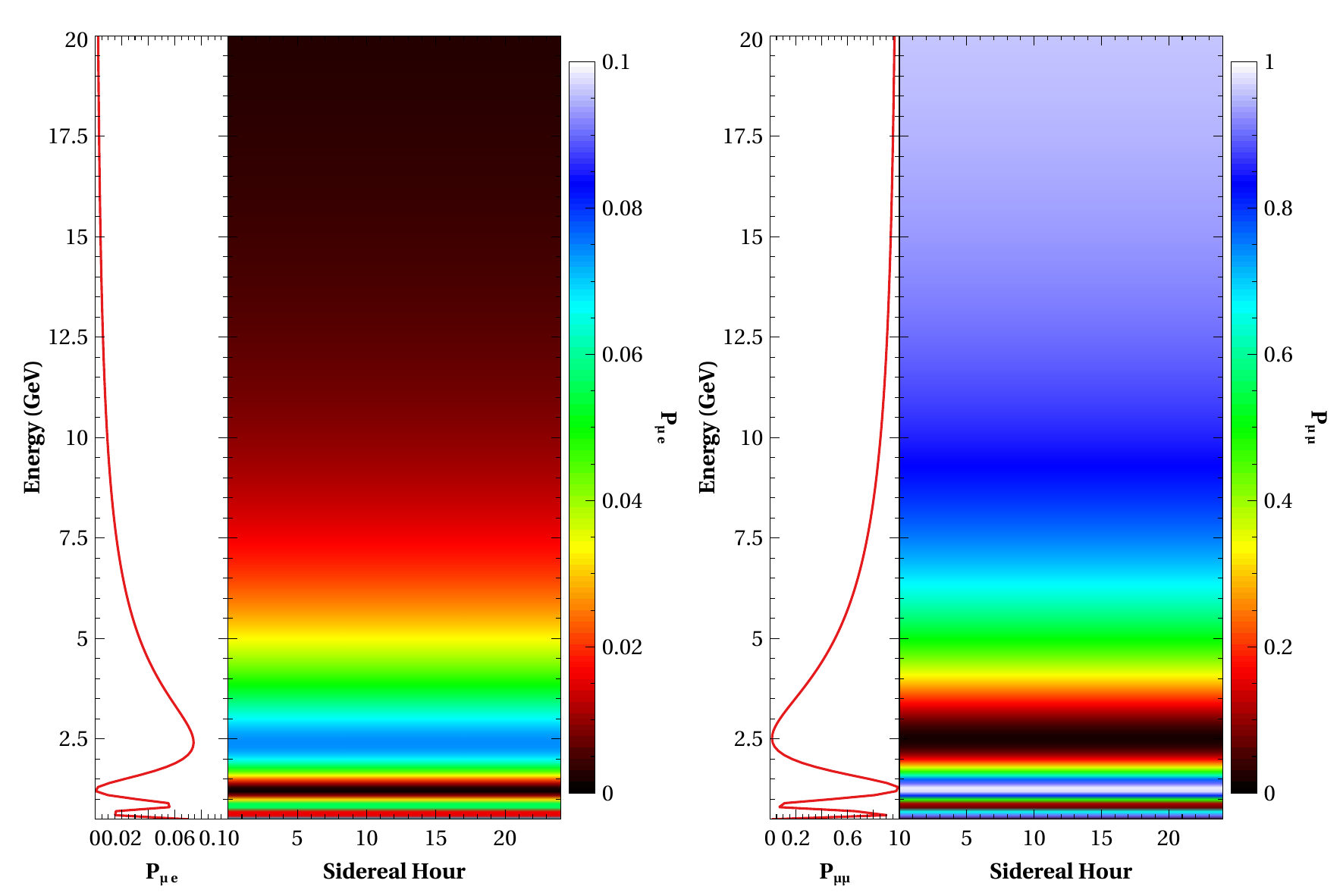}}
  \end{minipage}
  \hspace{0.2cm} 
  \begin{minipage}{0.48\textwidth}
     \centering
    \subfloat[Left: Appearance channel, Right: Disappearance channel in the SM and LIV case. ]{\includegraphics[width=\textwidth]{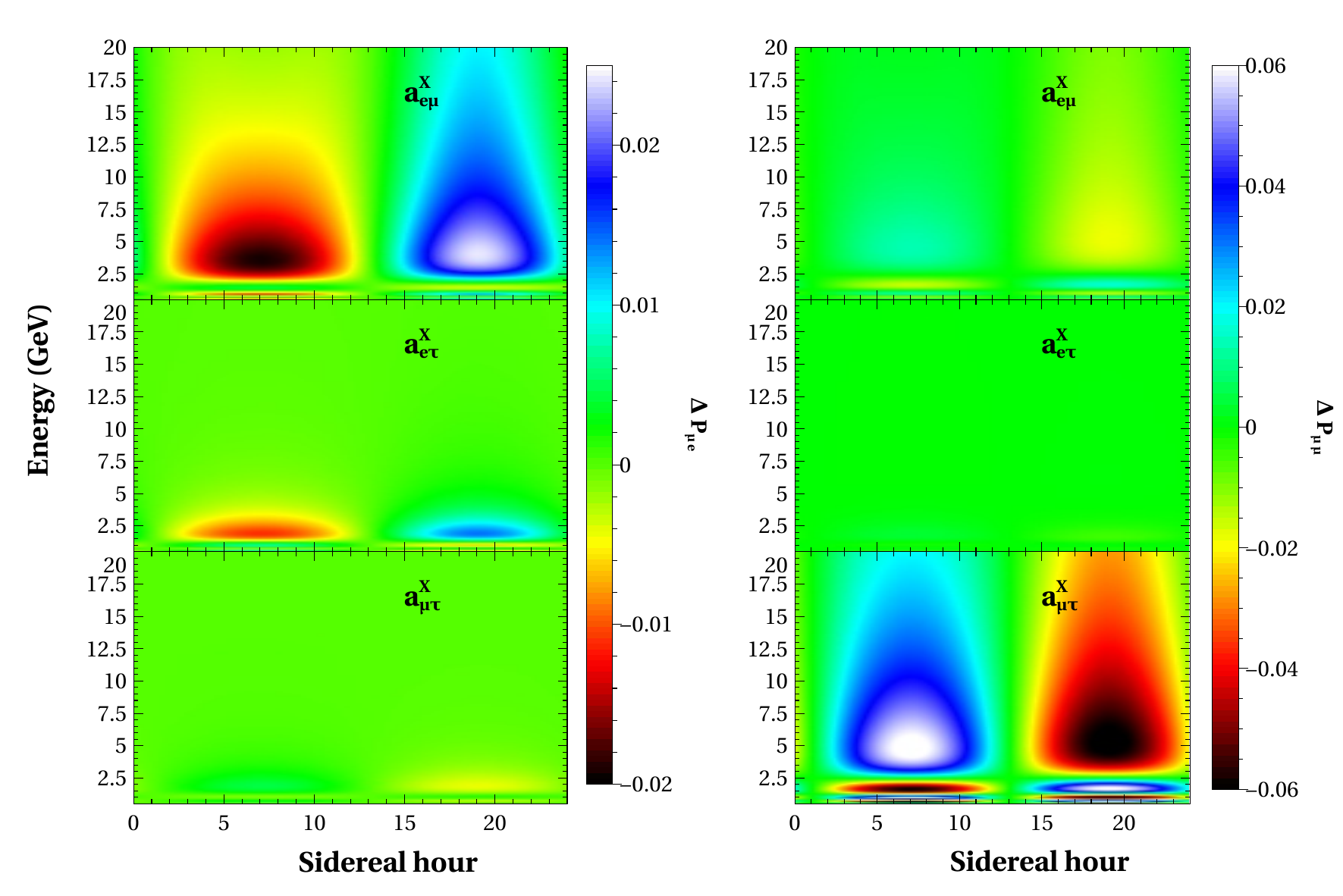}}
  \end{minipage}
  \begin{minipage}[b]{0.48\textwidth}
    \centering
    \subfloat[Left: Appearance channel, Right: Disappearance channel in the SM and LIV case.]{\includegraphics[width=\textwidth]{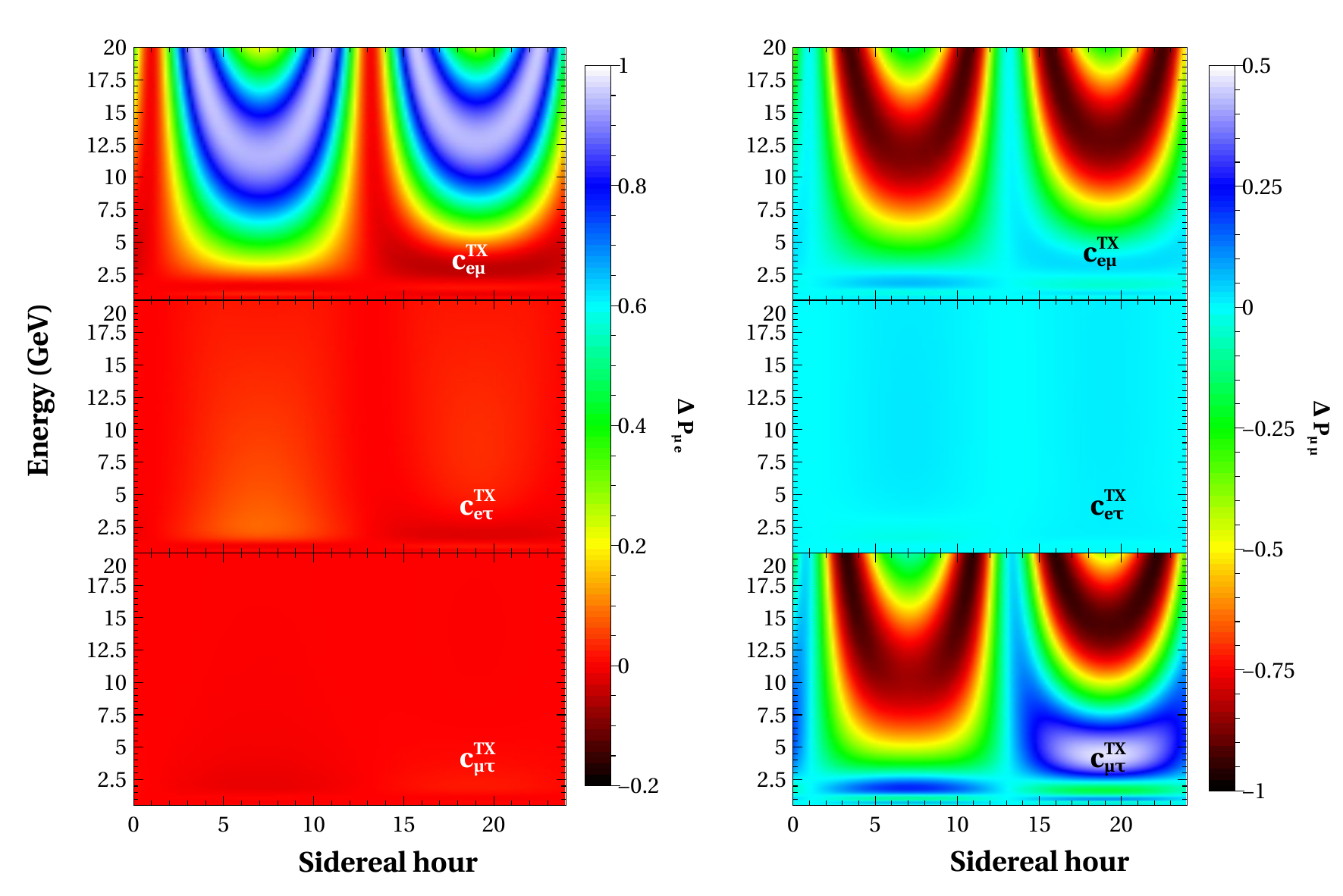}}
  \end{minipage}
  \hspace{0.2cm} 
  \begin{minipage}[b]{0.48\textwidth}
    \centering
    \subfloat[Left: Appearance channel, Right: Disappearance channel in the SM and LIV case.]{\includegraphics[width=\textwidth]{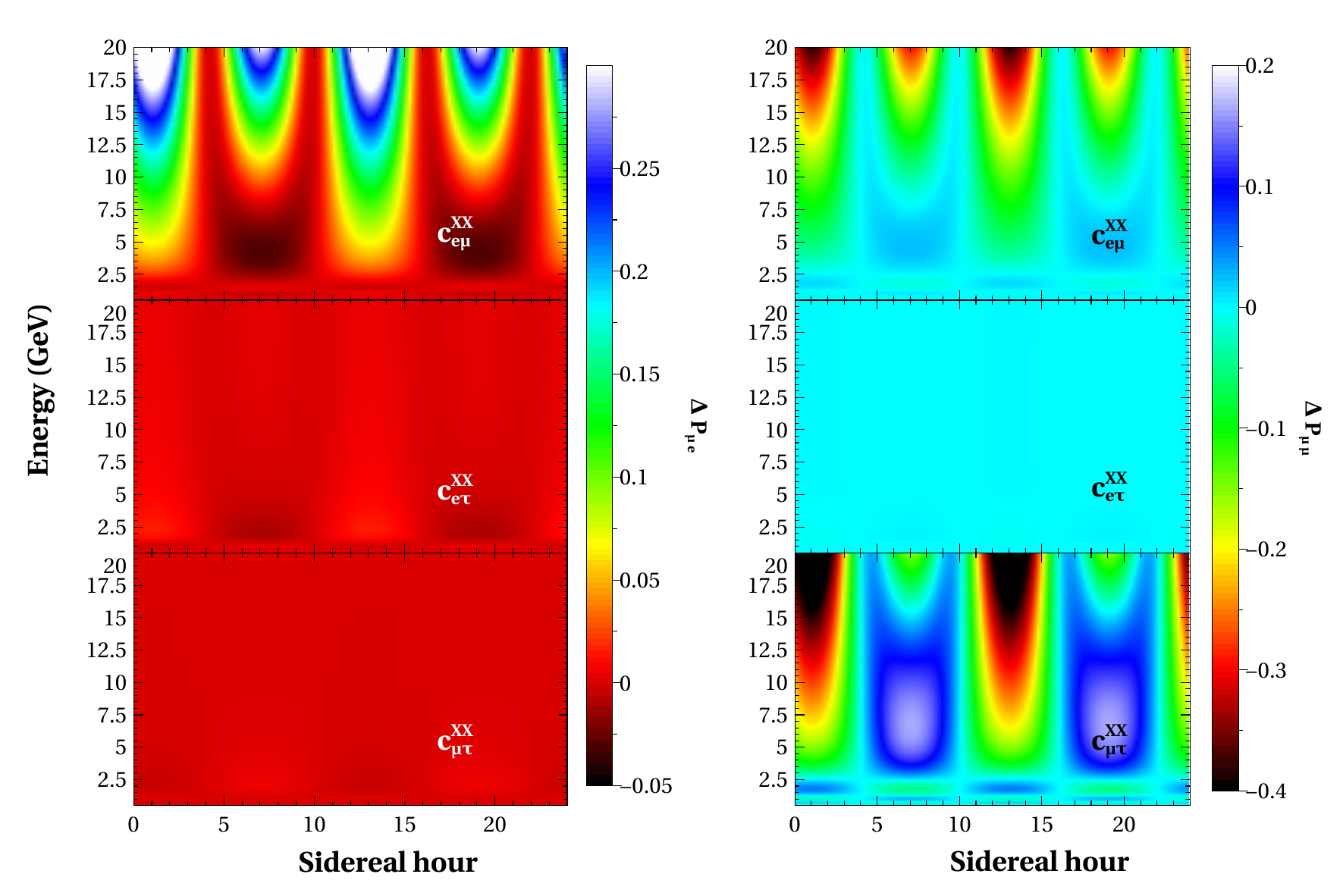}}
  \end{minipage}
  \begin{minipage}[b]{0.48\textwidth}
    \centering
    \subfloat[Left: Appearance channel, Right: Disappearance channel in the SM and LIV case.]{\includegraphics[width=\textwidth]{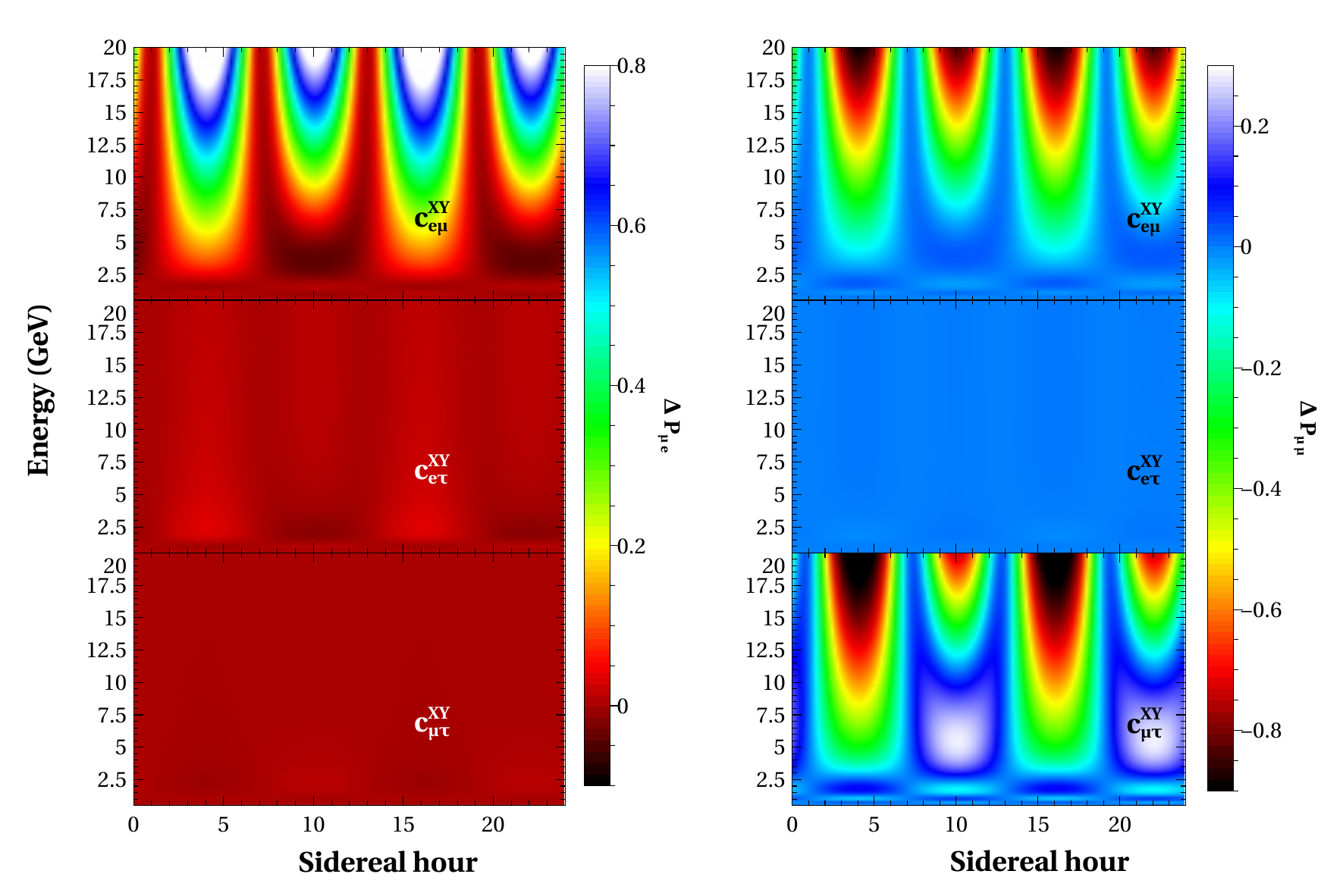}}
  \end{minipage}
  \hspace{0.2cm} 
  \begin{minipage}[b]{0.48\textwidth}
    \centering
    \subfloat[Left: Appearance channel, Right: Disappearance channel in the SM and LIV case.]{\includegraphics[width=\textwidth]{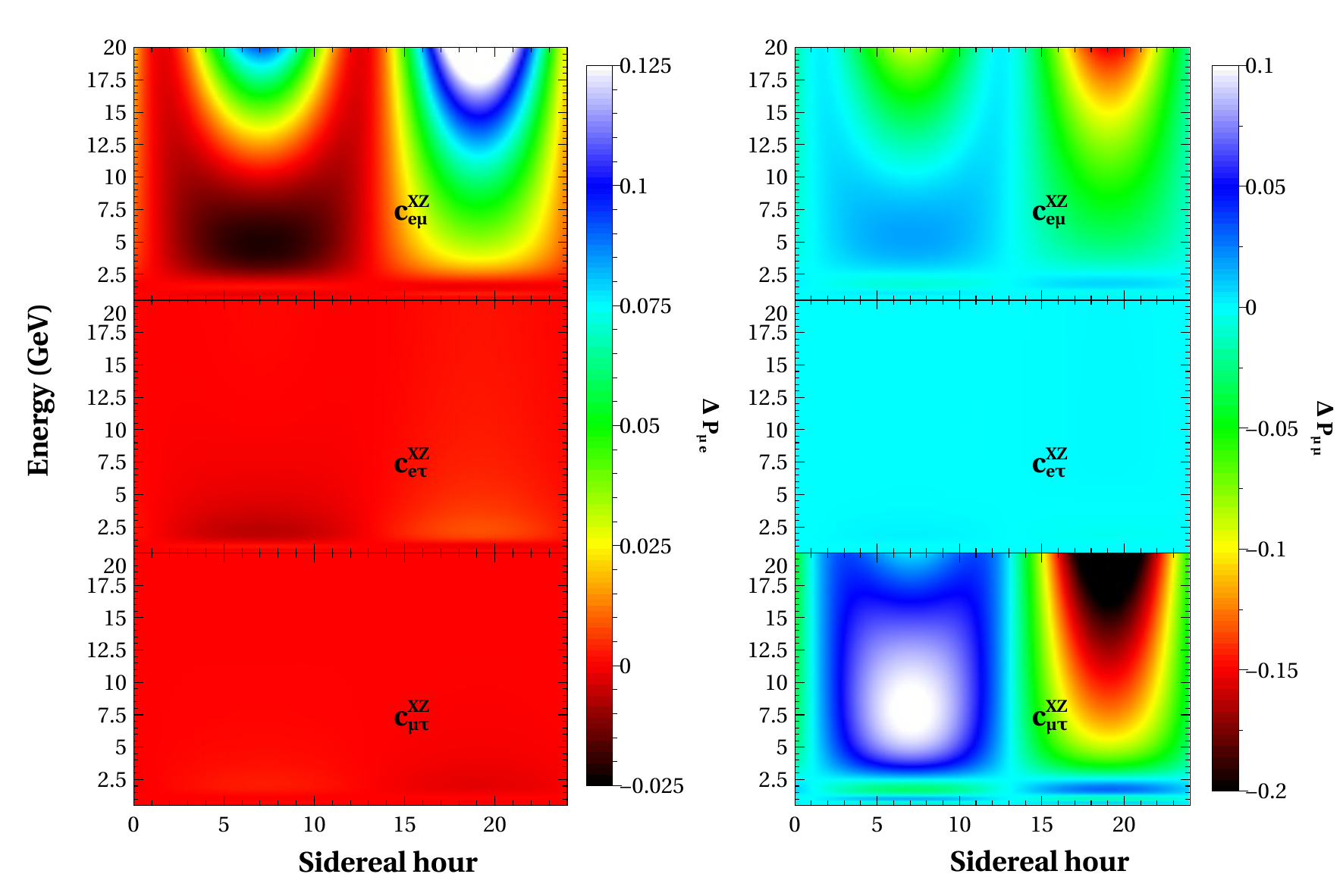}}
  \end{minipage}
  \caption{SM case: Panel(a) depicts the neutrino oscillation probability spectrum as a function of energy, alongside the probability distributions in terms of local sidereal time (LST) for the appearance (left) and disappearance (right) channels. These distributions are calculated using the oscillation parameters in Table~\ref{table3beamO}.
    SM + LIV Case: The panels (b) to (f) display the difference in probability distributions for the appearance  and disappearance channels for specific LIV parameter $a^{X}_{\alpha\beta}$, $c^{XY}_{\alpha\beta}$,, where $\alpha\beta$ represents $e\mu$, $e\tau$, and $\mu\tau$.  In each panel,  specific LIV parameter is set to $5 \times 10^{-23}$, while all others are zero.}
  \label{fig:widefigures}
\end{figure*}
The standard neutrino oscillation probability spectrum, based on the configuration of DUNE experiment, without considering LIV parameters, is shown in Fig.~\ref{fig:widefigures}(a). The left side of the Fig.~\ref{fig:widefigures}(a) illustrates the probability distribution for the appearance channel, whereas the right side presents the distribution for the disappearance channel. Both channels are represented the oscillation probability as functions of neutrino energy and local sidereal time (LST), highlighting how the oscillation probabilities evolve over these parameters in the absence of LIV effects. For the appearance channel, the first peak in oscillation probability for the DUNE  baseline is observed around 2.5 GeV. The low-energy flux distribution is specifically designed to emphasize this energy range. Within the SM framework, there are no significant effects observed along the sidereal time axis, as confirmed by the both panels of Fig.~\ref{fig:widefigures}(a). 
However, when LIV parameters are introduced, the smooth probability distribution becomes disrupted, resulting in significant distortions in the oscillation probability. These distortions depend on the specific LIV parameters along with energy and direction of neutrino, as clearly illustrated in Figs.~\ref{fig:widefigures}(b) through~\ref{fig:widefigures}(f). The presence of LIV parameters introduces energy-dependent modulations that alter the oscillation patterns, making these effects observable across different neutrino flux scenarios and energy ranges. The strength of each LIV parameter is fixed at $5 \times 10^{-23}$, a conservative value that is smaller than current experimental limits. 

At lower energies, specifically in the 1 to 5 GeV range, the a-type LIV parameters in $\emu$ and $\et$-channel significantly change the neutrino appearance channel probabilities, leading to noticeable deviations in oscillation behavior. On other hand, the $\mt$-type parameters have a dominant influence on the disappearance channel, where they prominently contribute in the leading terms of the probability formula, resulting in distinct modifications to the oscillation pattern. In the higher energy range of 5 to 20 GeV, $c$-type LIV parameters have a more substantial impact due to their energy dependence. As the energy increases, higher-order terms play a important role, leading to significant changes in the effects of individual LIV parameters. Above 15 GeV, the sidereal variation effect becomes particularly prominent for the $(c)^{TX}_{e\mu}$, $(c)^{XX}_{e\mu}$, $(c)^{XY}_{e\mu}$, and $(c)^{XZ}_{e\mu}$ parameter in the appearance channel, while the influence of $\et$ and $\mt$ parameters in this channel diminishes. In the disappearance channel, $\emu$-type parameters begin to show strong sidereal characteristics, which is less evident at lower energies. The sidereal variation strength of $\emu$ type parameters is considerably amplified at higher energies in both the appearance and disappearance channels. Meanwhile, $\mt$-type parameters show similar enhancement, but only in the disappearance channel. The $\et$-type parameters maintain moderate sidereal modulation strength at high energies in both channels.

\subsection{Sensitivity}
The broad energy spectrum of the DUNE experiment provides an opportunity to investigate how LIV parameters influence neutrino oscillation probabilities across a wide range of energies. Our analysis focuses on how these LIV parameters affect sidereal variation probabilities in different energy regimes, adapted to the flux scenarios. To evaluate sensitivity of DUNE experiment to detect sidereal effects and assess its maximal projected constrain for LIV parameters under null hypothesis, we use Poisson-likelihood chi-square statistics, as described in Appendix A. The best-fit value of oscillation parameters and their associated uncertainties are adopted from Ref~\cite{Esteban:2020cvm} for this work.   

In order to study the energy dependency of the sidereal variation pattern, it is convienet to analysis of neutrino oscillations piecewise over broad energy range. Therefore, the overall energy range from 0 to 20 GeV is divided into four distinct intervals: (1) 0-4 GeV, (2) 4-7.5 GeV, (3) 7.5-12 GeV, and (4) 12-20 GeV. This segmentation allows for a more detailed examination of the variations LIV parameters. To ensure a conservative approach in evaluating the influence of various LIV parameters across the specified energy intervals, all four distinct intervals are uniformly applied to the analysis of all parameter types. The strength of a LIV parameter also depends on its phase, which affects the sensitivity of an experiment to that specific parameter. Since the phases of these LIV parameters are unknown, a conservative approach involves marginalizing over the entire parameter space of the LIV phase ($\phi_{parameter}$). The uncertainties in the standard parameters $\Delta m^2_{31}$, $\theta_{23}$, and $\delta_{CP}$ can also significantly affect the sensitivity of experiment. Therefore, the test parameters $\theta_{23}$ and $\delta_{CP}$ are marginalized over the ranges $(41.0^\circ, 52.0^\circ)$ and $(0^\circ, 360.0^\circ)$, respectively. Test $\Delta m^2_{31}$ is also marginalized in both hierarchy.  The significance $\chi^{2}$ is plotted as a function of different parameters $\ax$, $\ctx$, $\cxx$, $\cxy$, and $\cxz$ (with $\alpha \beta = \emu, \et, \mt$) are presented from top to bottom, focusing on the low-energy flux scenario, in Fig~\ref{chi2::senA}. The panels are organized from left to right, representing the $e\mu$, $e\tau$, and $\mu\tau$ type parameters, respectively. Figure~\ref{chi2::senB} presents a similar format but with a tau-optimized flux scenario.
\begin{widetext}
  \begin{figure*}[htbp]
    \begin{minipage}{\textwidth}
    \centering 
    \includegraphics[height= 0.90\textwidth,width=0.90\textwidth]{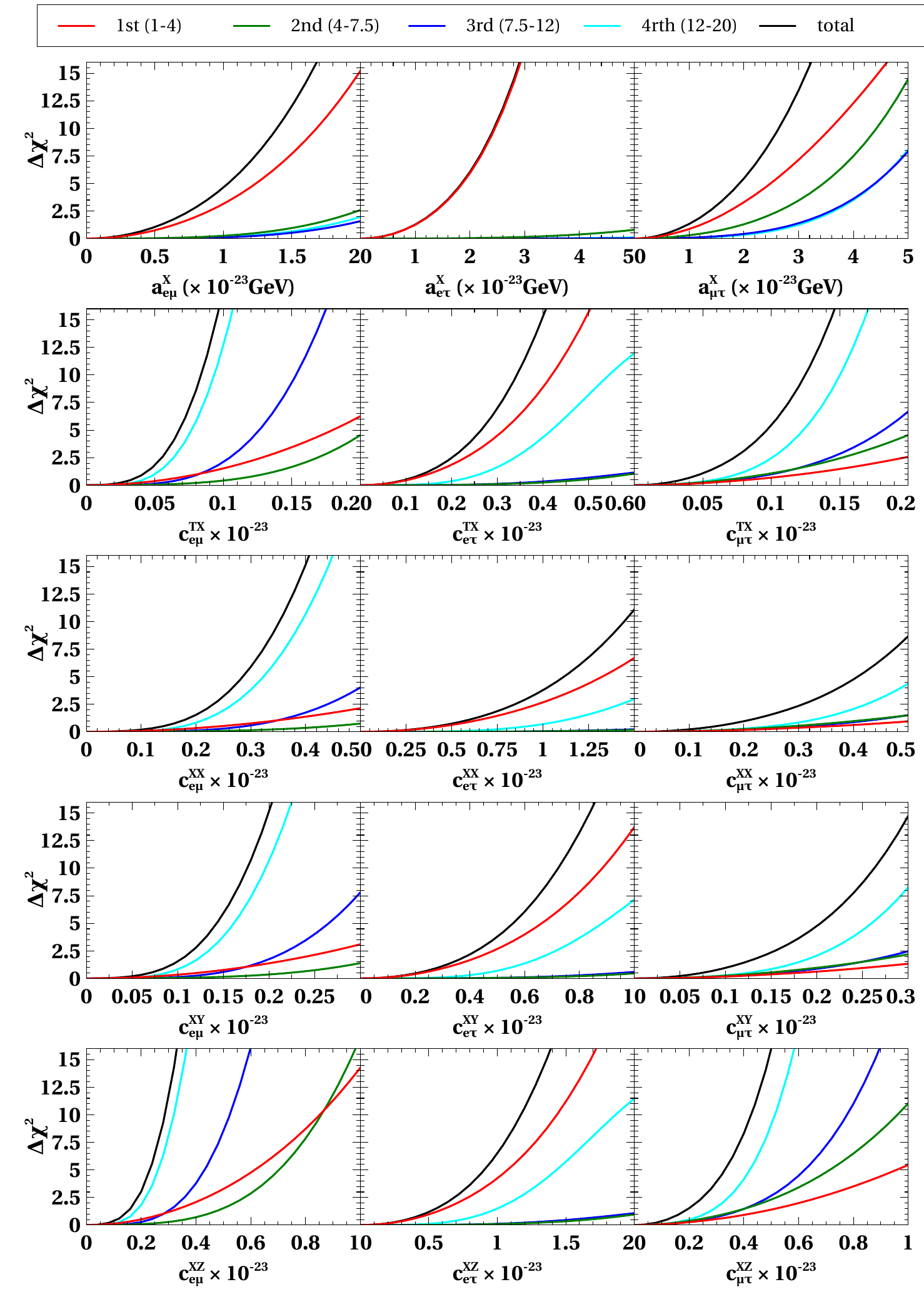}
    \end{minipage}
     \caption{(LE flux scenario) Sensitivity plots for the Lorentz invariance violation (LIV) parameters $\ax$, $\ctx$, $\cxx$, $\cxy$, and $\cxz$ are presented in sequence from top to bottom. The columns depict parameters categorized as $e \mu$, $e \tau$, and $\mu \tau$, arranged from left to right. Sensitivity across different energy bands is illustrated using various colored lines, while the black curve representing sensitivity over using events from entire energy range.}
     \label{chi2::senA}
  \end{figure*}

  \begin{figure}[htbp]
    \begin{minipage}{\textwidth}
    \centering 
    \includegraphics[height= 0.90\textwidth,width=0.90\textwidth]{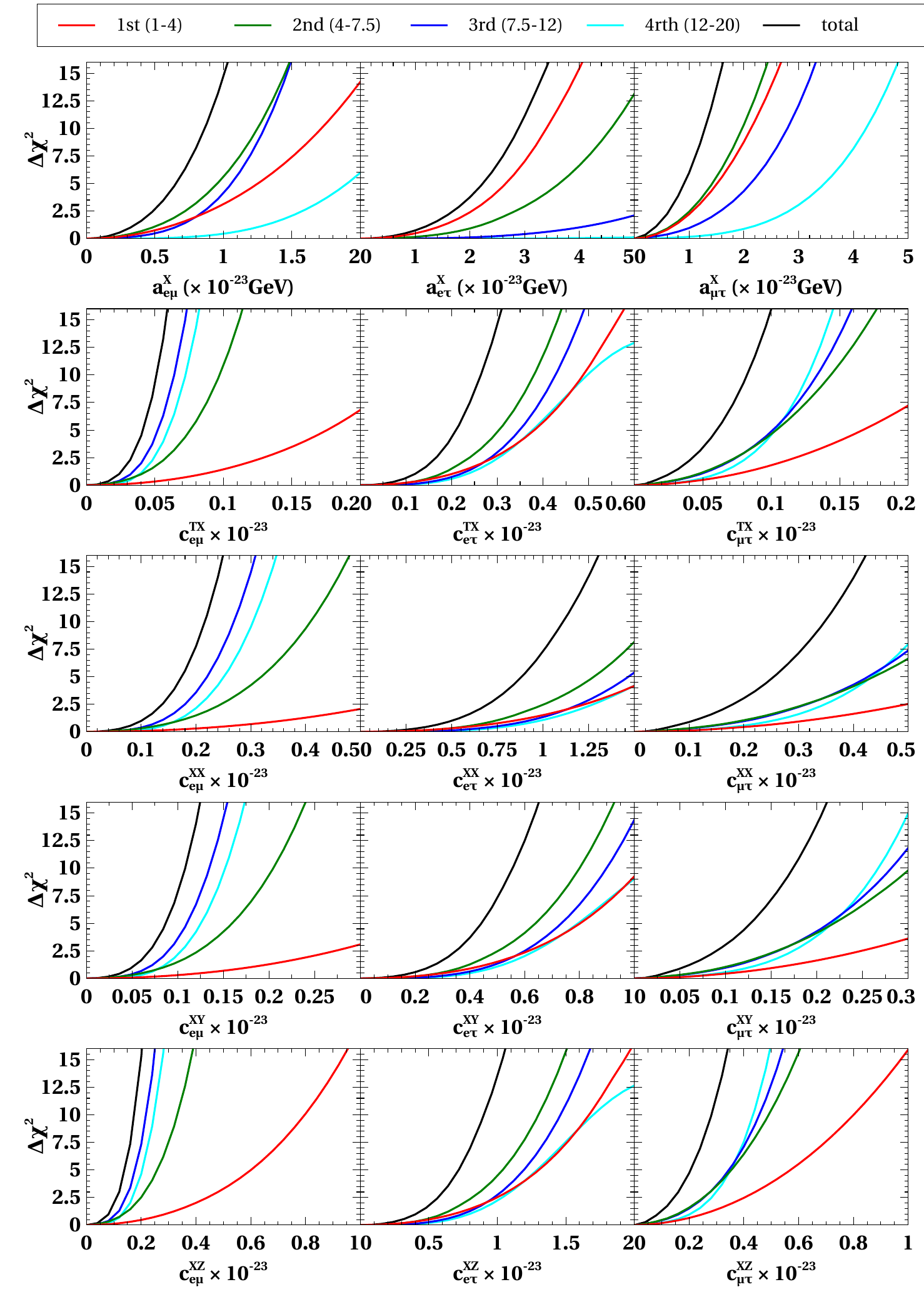}
    \end{minipage}
      \caption{Same as above but for tau-optimized flux.}
       \label{chi2::senB}
    \end{figure}   
   \end{widetext}
The directional and energy dependencies  of the LIV parameters are key parameters in the sensitivity curves.
The features of significance in Fig.~\ref{chi2::senA} for LE-flux scenario are as follows:
\begin{enumerate}
\item  The sensitivity of DUNE is primarily dominated by the first energy interval (0-4 GeV) for a-type parameters across all  $e \mu$, $e \tau$, and $\mu \tau$ channels ($a^{X}_{e\mu}$, $a^{X}_{e\tau}$, and $a^{X}_{\mu \tau}$). The second, third, and fourth energy bands contribute less significantly to the overall sensitivity,

\item  The high energy  interval (12-20 GeV) demonstrates the greatest sensitivity for c-type parameters in the $e \mu$, and $\mu \tau$ channels. For the $c^{TX}_{e\mu}$ parameter, the maximum sensitivity is observed in the fourth energy band, followed by the third, first, and second energy bands. This pattern is also observed for $c^{XX}_{e\mu}$, $c^{XY}_{e\mu}$, and $c^{XZ}_{e\mu}$. Similarly, for the $c^{TX}_{\mu \tau}$, $c^{XX}_{\mu \tau}$, $c^{XY}_{\mu \tau}$, and $c^{XZ}_{\mu \tau}$ parameters, the major contribution comes from the higher energy range ((12, 20) GeV). This is due to these parameters having a less significant impact at lower energies, making the larger energy intervals more important. The energy-dependent amplification of parameter strength shifts the modulation patterns towards higher energies.

\item   The sensitivity of the parameters $c^{TX}_{e\tau}$, $c^{XX}_{e\tau}$, $c^{XY}_{e\tau}$, and $c^{XZ}_{e\tau}$ is enhanced  with increasing energy, leading to notable variations in sidereal time at higher energy ranges. Since the LE flux is higher in the first energy band, these parameters exhibit their highest sensitivity in the first energy band, followed by the fourth band, with the second and third bands showing minimal sensitivity.
\end{enumerate}
It is to be noted that the overall sensitivity of the $e\tau$ parameters is lower as compared to the $e\mu$ and $\mu\tau$ parameters under the LE flux scenario.

In the tau-optimized fluxscenario, where the energy range extends from 1 GeV to 12 GeV, the flux distribution is relatively uniform across this range before gradually tapering off at higher energies. As a result, the contribution of the first energy interval (1-4 GeV) becomes less dominant compared to the higher energy interval. These higher energy interval exhibit more significant variations in oscillation probabilities, making them increasingly prominent in determining the overall sensitivity of the experiment. The main observations of Fig.~\ref{chi2::senA} for tau-optimized fluxscenario  are as follows:
 \begin{enumerate}
 \item The third energy interval (7.5-12 GeV) provides a more sensitive energy region for $c_{e\mu}$ parameters, while the second energy interval (4-7.5 GeV) is more sensitive to $ c_{e\tau}$ LIV parameters. Futhermore, $c_{e\mu}$ and $c_{\mu\tau}$ parameters exhibit notable sidereal modulation at higher energy ranges in appearance and disappearance channel respectively. As a result, parameters such as $c^{TX}_{e\mu}$, $c^{XX}_{e\mu}$, $c^{XY}_{e\mu}$, $c^{XZ}_{e\mu}$, $c^{TX}_{\mu\tau}$, $c^{XX}_{\mu\tau}$, $c^{XY}_{\mu\tau}$, and $c^{XZ}_{\mu\tau}$ are particularly sensitive in the higher energy intervals, making these regions crucial for detecting and constraining these parameters.

\item The sensitivity of $c_{\mu\tau}$ parameter receives comparable contributions from the second, third, and fourth energy intervals. This balanced contribution across multiple energy ranges indicates that no single energy interval dominates, and all play a significant role in determining the overall sensitivity for $c^{XY}_{\mu\tau}$.

\item The high energy interval does not significantly contrinbute to the sensitivity of a-type LIV parameters. The parameters $a^{X}_{\alpha \beta}$, which are not energy-dependent, are primarily influenced by the broad energy flux, with lower energy intervals contributing significantly to sensitivity. For instance, as illustrated in Fig.~\ref{fig:widefigures}(b), the probability difference for $a^{X}_{e \tau}$ is predominantly observed in the lower energy range, making the (1-4) GeV interval the most important for sensitivity. However, $a^{X}_{e\mu}$ and $a^{X}_{\mu \tau}$ parameters also exhibit small noticeable modulation at higher energy ranges, indicating that higher energy intervals do small contribute to overall sensitivity as well.
\end{enumerate}

 \begin{widetext}
 \begin{table*}[htbp]
     \caption{Summary of upper limits at 99.7\% C.L. that DUNE can set for all 27 LIV parameters under sidereal analysis using LE and HE flux scenarios.}
  \label{table4}
  \setlength{\tabcolsep}{13pt}
  \renewcommand{\arraystretch}{1.4}     
  \begin{center}
  \begin{tabular}{|p{2.8cm}|p{2.5cm}|p{2.5cm}|p{2.8cm}|p{2.8cm}|}
    \hline
              & Previous  & NOvA  & DUNE    & DUNE \\ 
    Parameter & Experimental & expected & LE Beam  & HE Beam \\ 
              &                             Limit~\cite{RevModPhys.83.11}&                      Limit~\cite{PhysRevD.109.075042}& ({\bf{This Work}}) & ({\bf{This Work}}) \\ \hline
    $|a^{X}_{e\mu}| = |a^{Y}_{e\mu}|$ & $  2.2 \times 10^{-20} \text{GeV} $& $  6.18 \times 10^{-23} \text{GeV} $ & $1.33 \times 10^{-23}$ GeV & $0.83 \times 10^{-23}$ GeV \\ \hline
    $|a^{X}_{e\tau}| = |a^{Y}_{e\tau}|$ & $ 4.4 \times 10^{-20}  \text{GeV} $&  $  9.64 \times 10^{-23} \text{GeV} $ & $2.34 \times 10^{-23}$ GeV & $2.78 \times 10^{-23}$ GeV \\ \hline
    $|a^{X}_{\mu\tau}| = |a^{Y}_{\mu\tau}|$ & $ 1.8 \times 10^{-23} \text{GeV} $&  $  6.75 \times 10^{-23} \text{GeV} $  & $2.51 \times 10^{-23}$ GeV & $1.23 \times 10^{-23}$ GeV \\ \hline

    $|c^{TX}_{e\mu}| = |c^{TY}_{e\mu}|$ &  $  9.0 \times 10^{-23} $&  $  1.32 \times 10^{-23} $ & $0.08 \times 10^{-23}$ & $0.05 \times 10^{-23}$ \\ \hline
    $|c^{TX}_{e\tau}| = |c^{TY}_{e\tau}|$ & $ 5.2 \times 10^{-18} $&  $  2.5 \times 10^{-23} $  & $0.33 \times 10^{-23}$ & $0.25 \times 10^{-23}$ \\ \hline
    $|c^{TX}_{\mu\tau}| = |c^{TY}_{\mu\tau}|$ & $ 3.7 \times 10^{-27} $& $  1.32 \times 10^{-23} $ & $0.12 \times 10^{-23}$ & $0.08 \times 10^{-23}$ \\ \hline

    $|c^{XX}_{e\mu}| = |c^{YY}_{e\mu}|$ & $ 4.5 \times 10^{-21} $&  $  7.57 \times 10^{-23} $  & $0.34 \times 10^{-23}$ & $0.21 \times 10^{-23}$ \\ \hline
    $|c^{XX}_{e\tau}| = |c^{YY}_{e\tau}|$ & $ 3.9 \times 10^{-17} $&  ...  & $1.39 \times 10^{-23}$ & $1.07 \times 10^{-23}$ \\ \hline
    $|c^{XX}_{\mu\tau}| = |c^{YY}_{\mu\tau}|$ & $ 2.5 \times 10^{-23} $& $  7.35 \times 10^{-23} $ & $0.51 \times 10^{-23}$ & $0.33 \times 10^{-23}$ \\ \hline

    $|c^{XZ}_{e\mu}| = |c^{YZ}_{e\mu}|$ & $ 1.1 \times 10^{-21}  $&  $  2.04 \times 10^{-23} $  & $0.28 \times 10^{-23}$ & $0.17 \times 10^{-23}$ \\ \hline
    $|c^{XZ}_{e\tau}| = |c^{YZ}_{e\tau}|$ & $ 1.8 \times 10^{-17} $&  $  3.72 \times 10^{-23} $  & $1.13 \times 10^{-23}$ & $0.87 \times 10^{-23}$ \\ \hline
    $|c^{XZ}_{\mu\tau}| = |c^{YZ}_{\mu\tau}|$ & $ 0.7 \times 10^{-23} $&  $  1.97 \times 10^{-23} $  & $0.41 \times 10^{-23}$ & $0.27 \times 10^{-23}$ \\ \hline

    $|c^{XY}_{e\mu}|$ & $ 2.2 \times  10^{-21} $&  $  3.78 \times 10^{-23} $  & $0.17 \times 10^{-23}$ & $0.10 \times 10^{-23}$ \\ \hline
    $|c^{XY}_{e\tau}|$ & $ 3.4 \times 10^{-18} $&  $  6.93 \times 10^{-23} $  & $0.70 \times 10^{-23}$ & $0.54 \times 10^{-23}$ \\ \hline
    $|c^{XY}_{\mu\tau}|$ & $ 1.2 \times 10^{-23} $&  $  3.67 \times 10^{-23} $  & $0.25\times 10^{-23}$ & $0.17 \times 10^{-23}$ \\ \hline
  \end{tabular}
  \end{center}
\end{table*}
 \end{widetext}
 Table~\ref{table4} presents the projected limits on non-isotropic non-diagonal LIV parameters derived from sidereal effects. Previous experimental limits on the non-istropic LIV parameters are taken from the table D31-D32 of ref~\cite{RevModPhys.83.11}. This work demonstrated that the DUNE experiment provides significant improvements in constraining all 27 LIV parameters, with the exception of  $c^{TX}_{\mu \tau}$ and  $c^{TY}_{\mu \tau}$. It is noted that the HE flux scenario offers more stringent constraints than the LE flux for a number of parameters, except for $c^{XZ}_{e\mu}$, $c^{YZ}_{e\mu}$, $c^{XZ}_{e\tau}$, $c^{YZ}_{e\tau}$, $c^{XZ}_{\mu\tau}$, and $c^{YZ}_{\mu\tau}$.  These results emphasize the importance of tau-optimized flux in improving the sensitivity of the DUNE experiment to LIV parameters. However, for certain parameters LE-flux flux is crucial. Therefore, a thorough senstivity assessment of all LIV parameters needs to take consideration of both LE and tau-optimized fluxscenarios to achieve a complete understanding of sensitivity and constraints.

%%%%%%%%%%%%
%%%%%%%%%%%%

\section{Summary}
\label{sec::summary}

This study explores sensitivity of non-isotropic LIV parameters by analyzing the sidereal effect within the framework of DUNE experiment. The oscillation probabilities and events are evaluated across different DUNE configurations using GLoBES simulations software with incorporation of LIV effect.
Our analysis demonstrates that LIV parameters show distinct behaviors across appearance and disappearance channels. The $e \mu$ and $e \tau$ type parameters primarily influence the appearance channels, while $\mu \tau$ type parameters have a stronger effect in the disappearance channels, as appearing at leading order.
The distortions noted in the probability distributions as a function of Local Sidereal Time (LST) and neutrino energy highlight the impact of sidereal frequency terms and directional factors incorporated within the LIV Hamiltonian. 
This study utilizes publicly accessible flux data to examine both LE and HE flux scenarios within the broad flux profile of the DUNE experiment. By analyzing these different energy regimes, this work aims to understand how variations in flux can influence the experimental sensitivity and constraints related to LIV parameters. The approach emphasizes the significance of these flux scenarios in providing a comprehensive understanding of capabilities of DUNE experiment,  in probing new constraints on LIV parameters with a $3\sigma$ confidence level under the null hypothesis. It is noted that, $c$-type parameters, which are energy dependent, show robust constraints when leveraging high-energy flux scenarios.

This work highlights the significant potential of the DUNE experiment to investigate non-isotropic LIV parameters. By utilizing intricate sidereal modulation and established flux scenarios, DUNE is uniquely equipped to extend the frontiers of current experimental constraints on LIV. This capability not only enhances the sensitivity of the experiment but also promises to greatly enrich our comprehension of LIV phenomena, paving the way for groundbreaking discoveries in fundamental physics.

\section{Acknowledgments}
We would like to thank Pedro A. N. Machado and Laura Fields for the suggestions related to publically available DUNE flux files. S. M., S. S. and V. S. are grateful for the financial support from the Department of Science and Technology (DST) in New Delhi, India, through their Umbrella Scheme for Research and Development. V. S. and L. S. also extend their thanks to the DST's FIST program, which supports the Department of Physics at CUSB. S. S. acknowledges the financial support provided by the Council of Scientific and Industrial Research (CSIR), New Delhi. Furthermore, L. S. is thankful for the funding received from the University Grants Commission under the Basic Scientific Research Faculty Fellowship Scheme (UGC-BSR), specifically through the Research Start-Up Grant (Contract No. F.30-584/2021 (BSR)).
%%%%%%%%%%%%%%%%%%%%%%%%%%%%%%%%%%%%%%%%%%%%%%%%%%%%%%%%%%%%%%%%%%%%
%%%%%%%%%%%%%%%%%%%%%%%%%%%%%%%%%%%%%%%%%%%%%%%%%%%%%%%%%%%%%%%%%%%%%

\section{Appendix}

\subsection{Chi Sqaure}
The sensitivity to Lorentz invariance violation (LIV) in the DUNE experiment is quantified using the $\Delta \chi^2$ statistic, defined by the formula:

\begin{widetext}

\begin{align}
\label{eq:chisq}
\Delta \chi^{2} = {\text{Min}} \Bigg[&2\sum_{x}^{\text{mode}}\sum_{j}^{\text{channel}}\sum_{i}^{\text{bin}}\Bigg\{
N_{ijx}^{\text{test}}(p^{\text{test}}) - N_{ijx}^{\text{true}}(p^{\text{true}})
+ N_{ijx}^{\text{true}}(p^{\text{true}}) \ln\frac{N_{ijx}^{\text{true}}(p^{\text{true}})}{N_{ijx}^{\text{test}}(p^{\text{test}})} \Bigg\}  \nonumber \\
\end{align}

\end{widetext}

In this formula, modes refer to neutrino and antineutrino types, channels represent both appearance and disappearance processes, and bins correspond to energy ranges, as specified in the DUNE Technical Design Report (TDR) files. This formula is employed to evaluate the discrepancy between the predicted neutrino event counts under a test hypothesis ($N_{ijx}^{\text{test}}(p^{\text{test}})$) and those under the true hypothesis ($N_{ijx}^{\text{true}}(p^{\text{true}})$). By minimizing this quantity, the sensitivity of the experiment to LIV effects is quantified. The summation over modes, channels, and bins ensures comprehensive inclusion of all relevant data dimensions. Through this approach, potential deviations from Lorentz invariance are effectively detected, thereby advancing the exploration of new physics beyond the Standard Model within the DUNE experiment.

\bibliography{LIVDUNE_Lakh_ref}

\end{document}